\documentclass{aa}

\begin{document}

   \thesaurus{10
(11.01.1; 
11.09.1; 
11.09.2; 
11.09.4; 
11.11.1; 
11.14.1; 
11.16.1; 
11.16.2; 
11.19.3; 
11.19.5)} 

   \title{THE MERGING SYSTEM AM 2049-691}


   \author {E.L.\,Ag\"uero\inst{1,2}, R.J.\,D\'{\i}az\inst{1},  
\& S.\,Paolantonio\inst{1}}


\institute{Observatorio Astron\'omico, Universidad Nacional de C\'ordoba, 
Laprida 854, 5000 C\'ordoba.//
email: aguero@oac.uncor.edu, diaz@oac.uncor.edu, santiago@oac. uncor.edu
\and
CONICET, Argentina}

   \date{Received June 10, 2000; accepted }

   \maketitle

   \begin{abstract}

	Spectroscopic and photometric observations of the peculiar object AM 2049-691 are presented here. Its systemic velocity is V$_{GSR}$ = (10956 $\pm$ 30 ) km s$^{-1}$, and the derived distance (H$_0$ = 75 km s$^{-1}$ Mpc$^{-1}$)  results 146 Mpc.  A bridge is observed between two very distinct nuclei whose separation is about 10 kpc, as well as two tails that emerge from the extremes SW and NE of the main body and extend up to 41 and 58 kpc respectively. The spectral characteristics of the all observed zones are typical of H II regions of low excitation. The internal reddening is quit high, particularly in the NE nucleus. All the derived equivalent widths of the H$\alpha$\,+\,[N II] lines indicate enhanced star formation compared with isolated galaxies, specially in the NE nucleus; the equivalent width corresponding to the integrated spectrum reflects starburst activity in the whole object, and is compatible with a merger of two disk galaxies. The observed characteristics of AM 2049-691 indicate it is a merger, where a overabundance of nitrogen is detected in one of the nuclei, which has the most evolved population and would be the most massive one. The detected total IR emission is not very high. The integrated total color B - V corresponds to a Sc-Scd galaxy and its average integrated population is about F7 type. Indicative B - V colors of the nuclei, corrected for internal absorption, are in agreement with the spectroscopic results. The central radial velocity dispersions at the nuclei suggest that the most massive galaxy would be the progenitor of the SW component. The observed radial velocity curve shows the presence of two subsystems, each one associated with a different nucleus. 

   \end{abstract}

%

\section{Introduction} \label{introduction}

The observation of the main characteristics of objects with double nucleus and/or distorted morphology indicative of evident gravitational interactions or mergers, allows to improve the knowledge about the involved physical processes. It makes possible to better understand the connections between interactions and different properties such as the separation and size of the components, spatial velocity distributions, global or local star formation activity, infrared emission, etc.

In an attempt to enlarge the available data of this kind of objects we present here spectroscopic and photometric observations of the little studied system AM 2049-691 (ESO 074-IG 020, IRAS 20494-6913) (Fig.~\ref{image}). Arp \& Madore (\cite{arp}) classified it as an interacting double object (Category 2) without a definitive statement about its probable merger condition. Lauberts and Valentijn (\cite{lauberts}) considered it is composed of a pair of galaxies of  Sb-c and Sb morphological types. 

A bridge is observed between two very distinct nuclei whose separation resulted of 14'', corresponding to a distance of 10 kpc (H$_0$ = 75 km s$^{-1}$ Mpc$^{-1}$). We also detected the bridge in H$\alpha$ emission. The system shows two tails; the tail that emerges from the SW extreme of the central body (0'.6 x 0'.2) is clearly visible up to $\approx$ 41 kpc from its center, whereas the other one reaches a distance of 58 kpc. At the tip of this last tidal tail it is seen what seems to be a dwarf galaxy as it was observed out of the debris of merging disk galaxies (Mirabel et al. \cite{mirabel}).

We have analyzed the main spectral characteristics of AM 2049-691, determined the principal excitation mechanisms, the physical conditions, abundances, the radial velocity distribution of the most important features, as well as its B magnitude and B - V, V - R, and R - I color indexes.


\section{Observations and Reductions} \label{observations}

\subsection{Spectra} \label{spectra}

Spectroscopic observations of AM 2049-691 were carried out on 1999 June 15 and 16 with the REOSC spectrograph coupled to the 2.15 m Ritchey-Chr\'etien telescope of The Complejo Astron\'omico El Leoncito, San Juan, Argentina, using a Tektronix 1024 x 1024 pixels CCD. The seeing was between 2'' and 3'' (FWHM).  Two set of spectra were obtained through a slit of $\approx$ 2'' wide and $\approx$ 3' long, along P.A. = 30$^{\circ}$.  One of  them  was taken using a 1200 lines mm$^{-1}$ grating over a wavelength range of $\approx$ 6400-7000 \AA, and the other one with a 300 lines mm$^{-1}$ grating covering the wavelength range of $\approx$ 3800-7200 \AA. The dispersions were 32 and 129 \AA mm$^{-1}$ and the resolutions 2.5 and 10 \AA, respectively. The spectra were corrected for atmospheric and Galactic extinction (A$_B$ = 0.15; Burstein \& Heiles \cite{burstein}), and flux calibrated with stars from the catalogue of Stone and Baldwin (\cite{stone}).

\subsection{CCD Photometry} \label{photometry}

Broadband B, V, R, and I photometric observations were performed on 1999 July 6, 7, and 8 with the CASLEO telescope and the CCD described above. The scale was 0''.27 pixel$^{-1}$. The seeing during observations was also 2''-3'' (FWHM). The obtained data were corrected for atmospheric extinction but they were not corrected for Galactic extinction. The photometric calibration was made using standard stars from Graham (\cite{graham}) observed through the same filters. 

Data reduction of spectra and images was accomplished using the standard methods in IRAF (developed by NOAO) reduction package. The journal of observations of AM 2049-691 is presented in Table ~\ref{log}.

\section{Spectroscopy} \label{spectroscopy}

In the spectra of the seven regions listed in Table ~\ref{fluxes} (regions 3 and 6 correspond to the centers of the NE and SW nuclear regions, respectively), obtained using the 300 lines mm$^{-1}$ grating, the lines measured were [O II] $\lambda$ 3727, H$\beta$, [O III] $\lambda$ 5007, [O I] $\lambda$ 6300, H$\alpha$, [N II] $\lambda$$\lambda$ 6548, 6584, and [S II] $\lambda$$\lambda$ 6717, 6731. The intensities were derived by fitting Gaussians to their profiles. The intensities of the [O III] $\lambda$ 4959 lines were constrained to the theoretical ratio [O III] $\lambda$ 4959 = 1/3 [O III] $\lambda$ 5007, so this line is not listed in Table ~\ref{fluxes}. The internal reddening correction was applied using the interstellar extinction curves given by Seaton (\cite{seaton}), assuming that the optical properties of the dust in AM 2049-691 are similar to those of the dust in the Galaxy. Intrinsic ratios H$\alpha$/H$\beta$ = 2.85 (Osterbrock \cite{osterbrock}) were adopted to derive the values of c, the logarithmic extinction at H$\beta$. 

For the mentioned seven regions the measured and corrected line intensities F$_{\lambda}$ and I$_{\lambda}$, relative to H$\beta$ = 1.00, as well as the errors which were estimated from the noise level around each line, are listed in Table ~\ref{fluxes}. The values of c and the corrected H$\beta$ fluxes are given at the bottom of this table. The spectra of all these regions present strong emission lines in the red zones (Fig.~\ref{spectra2}); their characteristics are typical of H II regions of low excitation, being the excitation considerably lower in region 3 than in region 6. The principal excitation mechanisms would be photoionization by young massive stars. The internal reddening is quite high, especially in region 3; a general decreasing trend is observed from NE to SW.  

\subsection{Abundances, physical conditions and equivalent widths} \label{abund}

The abundance ratios N(O)/N(H) and N(N)/N(H), the electron temperatures T$_e$ and densities N$_e$ were obtained for regions 1 to 7.  For the N(O)/N(H) abundances, the average values of N(O)/NH(H) derived from the empirical calibrations of Edmunds and Pagel (\cite{edmunds}) were adopted. The N(N)/N(H) abundances were derived by making the usual assumptions valid for HII regions. Expressions given by D\'{\i}az (\cite{diaz}) were used for the involved ionic abundances. The electron temperatures were obtained by searching the required values of T$_e$ for the adopted N(O)/N(H) abundances; the electron densities were derived from the [S II] $\lambda$6717/$\lambda$6731 ratios (\cite{osterbrock}). The results are presented in Table ~\ref{abundances}. The electron temperatures are rather low, but they are in the range of normal values for H II regions; electron densities are also within that range.

The derived nitrogen and oxygen abundances present two maxima corresponding to the region 3 (NE nucleus) and region 6 (SW nucleus), being the abundances of both elements higher in the NE nucleus than in the SW one. In region 3 the N(O)/N(H) and N(N)/N(H) abundance ratios are 2.0 and 1.2 times of the corresponding solar abundances. In region 6 both N(O)/N(H) and N(N)/N(H) ratios are about 1.1 of the respective solar values. The N(N)/N(O) ratios in the region 3 and towards the NE, are practically coincident with those of the galactic emission regions (Shaver et al. \cite{shaver}), indicating the same proportions of the involved elements. Towards the SW these ratios increase, being in region 6 about twice of those of galactic regions; this indicates a comparative overabundance of N with respect to the O, which is reflected in the relatively high [N II] $\lambda$ 6584/H$\alpha$ ratios. If this excess is due to an enhancement of nitrogen abundance after a succession of short bursts (Contini et al. \cite{contini}), that overabundance suggests that the SW nucleus has undergone previous star formation bursts. Overabundance of N was also detected in the Seyfert component of the interacting pair of galaxies NGC 5953 (Seyfert nucleus) and NGC 5954 (LINER) (Gonz\'alez Delgado \& P\'erez \cite{gonzalez}). 

The equivalent widths EW(H$\alpha$ + [N II]) also show two maxima at the two nuclei (Fig.~\ref{ewidth}), being EW(H$\alpha$ + [N II]) = 67\,\AA\, and 48\,\AA\, for the NE and SW nuclei respectively. All the obtained values indicate enhanced star formation activity compared with isolated galaxies, especially in the NE nucleus. The equivalent width EW(H$\alpha$ + [N II]) = 58 \AA\, derived from the integrated spectrum of AM 2049-691 reflects there is star formation activity in the whole object, which could be favored with the usually large amounts of gas that spiral galaxies have, and is compatible with a merger of two disk galaxies (Liu \& Kennicutt \cite{liu}). In this system the star formation activity, presumably induced by the interactions, takes place in both nuclei being more significant in the north-eastern one, as detected in some other pairs (Sekiguchi \& Wolstencroft \cite{sekiguchi}), and in the whole object. This differs from the results of Joseph et al. (\cite{joseph}) who found evidence of this activity in only one member of their observed pairs. 

The H$\alpha$ equivalent widths determined for the NE and SW nuclei are EW(H$\alpha$) = 44 and 25 \AA; close ages of $\approx$ 9 x 10$^6$ yr are derived for their bursts of star formation according to the standard model for instantaneous bursts with metallicities of about 2 Z$_{\odot}$ and 1 Z$_{\odot}$ (Leitherer \& Heckman \cite{leitherer}) respectively.

The O and N abundances, N(N)/N(O) ratios, electron temperatures and densities, internal reddenings, and equivalent widths are different in the NE and SW nuclei, reflecting the different evolutions they have undergone. 

\subsection{Radial Velocities} \label{velocity}

Radial velocities were derived from the spectrum obtained with the 1200 lines mm$^{-1}$ grating by measuring the centroids of Gaussian curves fitted at the profiles of the strongest emission lines. The resulting heliocentric radial velocities of NE and SW nuclei are V$_{NE}$ = (10977 $\pm$ 18) km s$^{-1}$ and V$_{SW}$ = (11144 $\pm$ 13) km s$^{-1}$ respectively. The average velocity was adopted as the systemic velocity of AM 2049-691, which referred to the Galactic System of Rest is V$_{GSR}$ = (10956 $\pm$ 30 ) km s$^{-1}$, and the derived distance (H$_0$ = 75 km s$^{-1}$ Mpc$^{-1}$)  results 146 Mpc.  AM 2049-691 is included in the survey of Sekiguchi \& Wolstencroft (\cite{sekiguchi}) who reported both nuclear velocities (with errors larger than ours) without any detailed kinematical analysis; the values obtained here are consistent  with theirs.

Slight asymmetries were detected in the nuclear emission lines and they can be fitted by secondary components, 100 km s$^{-1}$ blueward at SW nucleus and 150 km/s redward at NE nucleus.

The Na I ((5893 \AA) absorption line was also detected in the continuum emission of each nuclei, with velocities V$_{NE}$ = (10977 $\pm$ 20) km s$^{-1}$ and V$_{SW}$ = (11140 $\pm$ 20) km s$^{-1}$. This absorption line appeared with almost the same equivalent width ($\approx$ 2 \AA) at both nuclei but the FWHM of each line were different. Their measured  radial velocity dispersions were $\sigma_{NE}$ = (280 $\pm$ 20) km s$^{-1}$  and  $\sigma_{SW}$ = (330 $\pm$ 20) km s$^{-1}$ and the deconvolved central radial velocity dispersion of the stellar systems are $\sigma_{NE}$ = (225 $\pm$ 20) km s$^{-1}$  and  $\sigma_{SW}$ = (280 $\pm$ 20) km s$^{-1}$. As the nuclear dynamics of ellipticals and normal bulges in spirals has been found to be indistinguishable (Kormendy \& Illingworth \cite{kormendy}) we can consider, at first approximation, the central velocity dispersions as indicative of the relative masses of the original systems. Thus the progenitor galaxy of the SW component would be the most massive one. The stellar radial velocity dispersion allows us to estimate the mass and tidal radius (e.g. Bowers \& Deeming \cite{bowers}) of each nuclear-bulge component, which turns out to be roughly 5.9 x 10$^{10}$ M$_{\odot}$  and 6.1 kpc for NE nucleus; and  9.4 x 10$^{10}$ M$_{\odot}$  and 8.4 kpc for SW one.

Following the results of Kormendy \& Illingworth (\cite{kormendy83}) about the L\,$\approx$\,$\sigma$$^n$ relation for disk-galaxy bulges, the values presented here are roughly consistent with the progenitor systems being spiral galaxies with M$_B$ $\approx$ -21.  These values are in accordance with the global photometric properties presented in Section ~\ref{magnitudes} and the spectrophotometric results discussed in Section ~\ref{abund}.  

The emission lines velocity distribution (along P.A. = 30$^{\circ}$) is illustrated in Figure ~\ref{radvel}a, where the open circles correspond to the two distinct nuclei. The radial velocity curve along the line joining both nuclei shows the presence of two different components separated by a velocity discontinuity of $\approx$ 100 km s$^{-1}$, and a first glance of the curve suggests that each one is associated with a different nucleus. The NE component has an approximate solid body (SB) behavior at all the measured positions and the SW component appears to have a strong asymmetry in the velocity values respect to the nucleus. However, a close inspection of the spectra showed us that the H$\alpha$ emission has a minimum between both nuclei at 1/2 of the distance from NE to SW nucleus. Then the three points after the velocity discontinuity, have photometric continuity with the NE emission complex, as shown in Figure ~\ref{radvel}b. As the tidal radius of NE system is smaller (r$_T$\,$\approx$ 6 kpc), this feature could be caused by tidal disruption of the NE gaseous system, part of which could have became gravitationally bounded to the SW body, apparently the most massive one. This peculiar kinematic feature and the strong H$\alpha$ emission makes this merging galaxy an ideal target for two-dimensional spectroscopy.

The global appearance of the rotation curve is solid body (SB) like at 70\% of the observed positions. SB rotation curves appear more frequently in low luminosity galaxies and in interacting disk galaxies (Keel \cite{keel}). In the case of AM 2049-691, the SB appearance would not necessary correspond to a spherical halo mass distribution, since recent numerical simulations have shown that an appropriate combination of perturbation and dust obscuration in the disk can explain the SB appearance of an interacting galaxy rotation curve at a wide range of radii (D\'{\i}az et al. \cite{diaz00}).

As a whole system, AM 2049-691 shows a velocity amplitude of $\approx$ 330 km s$^{-1}$ (sin i) $^{-1}$ within a diameter of $\approx$ 23 kpc and the kinematical center is possibly located on the line joining both nuclei. The total keplerian mass inside a radius of 11.5 kpc is $\approx$ 1.4 x 10$^{11}$ M$_{\odot}$ ; as the orientation is unknown and this system is far from relaxation, this is only a very rough estimate, but consistent with the luminosities reported in the next section.

\section{Photometry} \label{photometry2}

\subsection{Infrared Data} \label{infrared}

AM 2049-591 is not a luminous infrared object (L$_{FIR}$ less than 10$^{11}$ L$_{\odot}$) but it has an appreciable IR emission. Its far infrared flux, FIR = 1.26 x 10$^{-11}$(2.58 S$_{60}$\,+\, S$_{100}$) (Londsdale et al. \cite{londsdale}) = 7.4 x 10$^{-11}$ erg cm$^{-2}$ s$^{-1}$, calculated using the appropriated data from the IRAS Point Source Catalog (\cite{iras}) leads, adopting a mean distance of 146 Mpc (H$_0$ = 75 km s$^{-1}$ Mpc$^{-1}$), to the IR luminosity L$_{FIR}$ = 5 x 10$^{10}$ L$_{\odot}$ which is compatible with a merger system. Its comparatively low infrared colors $\alpha$(60, 25) = -2.5 and $\alpha$(100, 60) = -1.5 indicate it is a nonactive object, which is coherent with the spectroscopic results.

In this system, the interaction seems not to have produced very high IR emission when compared with the IR luminosity of typical mergers (L$_{FIR}\,\approx$ 5 x 10$^{11}$ L$_{\odot}$).

\subsection{Magnitudes and Colors} \label{magnitudes}

B magnitudes and B - V, V - R, and R - I colors of AM 2049-691 were derived using circular apertures with increasing radii (after removing the stars) centered on a point equidistant from the two nuclei. The photometric useful frames were smaller than the total estimated system size, so asymptotical extrapolations of the obtained values were used to estimate the total magnitudes. The obtained results are B = 14.40,  B - V = 0.52, V - R = 0.47, and R - I = 0.59.  The uncertainties are $\pm$0.02 in B, $\pm$0.05 in B - V, and $\pm$0.06 in V - R and R - I.  The magnitudes B and R obtained here are coherent with those of Lauberts \& Valentijn (\cite{lauberts}), considering as the total magnitude of the whole system that derived from the sum of their individual values for the NE and SW components. The integrated total color  B - V would correspond to a Sc-Scd galaxy (Roberts \& Haynes \cite{roberts}) and indicates that the integrated population is about F7 type.

Indicative magnitudes and colors of both nuclei were derived. Diaphragms with radii of 5'' were used for both nuclei. The B magnitudes and B - V colors corresponding to the northeastern and southwestern nuclei are 16.28 and 16.66, and 0.75 and 0.87 respectively. After correcting these values for internal absorption by adopting A$_{\lambda}$ = E$_{B-V}$.X(x) and the extinction curves given by Seaton (\cite{seaton}), the B - V colors for the NE and SW nuclei became - 0.25 and 0.19, corresponding respectively to average integrated populations of about B1 and A6 types. These values clearly indicate that the two nuclei are star forming regions, being the observed star formation activity more intense in the NE nucleus than in the other one, as found from the spectroscopic data. 


\section{Summary and Conclusions} \label{summary}

We performed CCD spectroscopic and broadband B, V, R, and I photometric observations of AM 2049-691 that is a pair of comparably sized interacting galaxies of morphological types Sb-c and Sb with a separation comparable to their sizes. From the derived information the principal results are:

The spectral characteristics of the all studied regions are typical of H II regions of low excitation; their dominant excitation mechanisms would be the photoionization by young massive stars. The internal reddening is quite high, especially in the northeastern nucleus, and reveals an inhomogeneous obscuration.

All the derived equivalent widths of the H$\alpha$ + [N II] lines indicate enhanced star formation activity compared with isolated galaxies, being this activity more intense in the NE nucleus. The equivalent width corresponding to the integrated spectrum suggests starburst activity in the whole object, and is compatible with a merger of two disk galaxies. For both nuclei the derived starburst ages is about 9 x 10$^6$ yr suggesting that the merger process triggered the present star formation bursts in both progenitor galaxies at the same time.

The N(N)/N(O) ratio suggests in NE nucleus the same proportion of oxygen and nitrogen as in galactic emission regions; at the SW nucleus this ratio is about twice of those values, indicating there a comparative overabundance of N with respect to O, which is reflected in the relatively high [N II] $\lambda$6584/H$\alpha$ ratios. AM 2049-691 is a merger where overabundance of nitrogen is detected in one of the nuclei, which has the most evolved population and would be the most massive one.

AM 2049-691 is not a very luminous infrared system but has an appreciable IR emission: L$_{FIR}$ = 5 x 10$^{10}$ L$_{\odot}$; its comparatively low far infrared colors $\alpha$(60,25) and $\alpha$(100,60) indicate it is a nonactive object, which is consistent with the derived spectroscopic data.

The integrated total color B - V corresponds to a Sc-Scd galaxy and its average integrated population would be about F7 type. Indicative B - V colors of the nuclei, after correcting for internal extinction, suggest they are regions of star formation activity, specially the NE nucleus as found from the spectroscopic observations.

The central radial velocity dispersions at the nuclei indicate that the most massive galaxy was the progenitor of the SW component. The observed radial velocity curve shows the presence of two components (each one associated to a different nucleus) that undoubtedly correspond to the merging galaxies, what is confirmed by the distinct spectrophotometric and photometric properties shown by the structures associated to each nucleus.

\vspace{0.5cm}
AM 2049-691 is an ongoing merger with a spatially extended star formation activity. It has a high level of disruption and interpenetration and shows a double set of morphological, spectrophotometric and kinematical subsystems.

\begin{acknowledgements}

We acknowledge the colaboration of M.\,Campos at the observing run.

\end{acknowledgements}

\newpage
\begin{center}

\end{center}
\vskip 0.4cm

\begin{figure}
\caption [f1.ps] { Blue image (from ESO plates) of AM 2049-691. North is at the top and east to the left.  The lines mark P.A. = 30$^{\circ}$. \label{image}}
\end{figure}

\begin{figure}
\caption [f2.ps] { Spectrum of: (a) region 3 (NE nucleus), (b) region 6 (SW nucleus).  Dispersion  is 129 \AA  mm$^{-1}$. \label{spectra2}} 
\end{figure}

\begin{figure}
\caption [f3.eps] { Equivalent widths vs. distance.  Empty circles correspond to the centers of the northeastern and southwestern nuclear regions. Distances are given with respect to region 3, being positive to the northeast. \label{ewidth}}
\end{figure}

\begin{figure}
\caption [f4.ps] {(a) Radial velocity distribution along P.A. = 30º. The values correspond to the weighted average velocities from the H$\alpha$, [N II] $\lambda$$\lambda$ 6548, 6584 lines. Bars indicate errors. Symbols and distances are as in Fig.~\ref{ewidth}. 
	  (b) H$\alpha$ emission profile along P.A. = 30$^{\circ}$. \label{radvel}}
\end{figure}

\begin{table} 
\footnotesize
\caption{JOURNAL OF AM 2049-691 OBSERVATIONS} \label{log}
\begin{tabular} {lccc}
\hline
{\bf Date} & {\bf Spectral Region} & {\bf Exposure Time} & {\bf Comments} \\ 
\vspace{0.2cm}
{}  &  {\bf or Filter} & {\bf (seconds)}  & {} \\
\hline
1999 June 15 & $\lambda\lambda$ 6400-7000 & 1200 & P.A.=30$^{\circ}$ \\
1999 June 15 & $\lambda\lambda$ 6400-7000 & 1200 & P.A.=30$^{\circ}$ \\
1999 June 16 & $\lambda\lambda$ 3800-7200 & 1800 & P.A.=30$^{\circ}$ \\
1999 June 16 & $\lambda\lambda$ 3800-7200 & 1800 & P.A.=30$^{\circ}$ \\
1999 July 6 & B & 30 & \\
1999 July 6 & B & 30 & \\
1999 July 6 & B & 300 & \\
1999 July 6 & B & 300 & \\
1999 July 6 & B & 300 & \\
1999 July 6 & V & 30 & \\
1999 July 6 & V & 30 & \\
1999 July 6 & V & 300 & \\
1999 July 6 & V & 300 & \\
1999 July 7 & R &  60 & \\
1999 July 7 & R &  60 & \\
1999 July 7 & I &  60 & \\
1999 July 7 & I &  60 & \\
\hline

\end{tabular}
\end{table}

\begin{table} 
\footnotesize
\caption{LINE INTENSITIES RELATIVE TO H$\beta$} \label{fluxes}
\begin{tabular}{lcccccccc}
\hline
{} & {} & {} & {} & {}  & {\bf F$\lambda$/F$\beta$} & {} & {} & {}\\ 

{\bf Ion} & {\bf $\lambda$} & {} & {} & {}  & {\bf I$\lambda$/I$\beta$} & {} & {} & {}\\ 

{} & {\bf (\AA)} & {} & {} & {}  & {\bf $\sigma$} & {} & {} & {}\\ 

\hline

{} & {} & {Region 1} & {Region 2}  
& {Region 3}  & {Region 4} & {Region 5} & {Region 6} & {Region 7}\\ 
\vspace{0.2cm}
{} & {} & {5.1''} & {2.0''}  & {0''}  & {-2''.0} & {-12''.2} & {-14''.2} & {-16''.3}\\ 
\hline

       &    & 1.18 & 0.89 & 0.60 & 0.76 & 1.63 & 1.10 & 1.17 \\
$[O II]$ & 3727 & 3.11 & 2.10 & 1.61 & 1.72 & 3.12 & 2.16 & 2.33 \\
\vspace{0.2cm}
       &    & 0.02 & 0.01 & 0.01 & 0.01 & 0.09 & 0.06 & 0.06 \\
   &    & 1.00 & 1.00 & 1.00 & 1.00 & 1.00 & 1.00 & 1.00 \\
$H\beta$ & 4861 & 1.00 & 1.00 & 1.00 & 1.00 & 1.00 & 1.00 & 1.00 \\
\vspace{0.2cm}
   &    & -- & -- & -- & -- & -- & -- & -- \\
   &    & 0.60 & 0.29 & 0.21 & 0.56 & 0.60 & 0.74 & 0.88 \\
$[O III]$ & 5007 & 0.52 & 0.26 & 0.18 & 0.50 & 0.55 & 0.68 & 0.80 \\
\vspace{0.2cm}
    &     & 0.01 & 0.01 & 0.01 & 0.01 & 0.01 & 0.01 & 0.02 \\
   &    & 0.31 & 0.26 & 0.19 & 0.24 & 0.12 & 0.17 & 0.38 \\
$[O I]$ & 6300 & 0.11 & 0.10 & 0.06 & 0.10 & 0.06 & 0.08 & 0.18 \\
\vspace{0.2cm}
   &    & 0.01 & 0.01 & 0.01 & 0.01 & 0.01 & 0.01 & 0.01 \\
   &    & 1.31 & 1.10 & 1.24 & 1.59 & 1.40 & 2.08 & 2.39 \\
$[N II]$ & 6548 & 0.39 & 0.37 & 0.36 & 0.57 & 0.62 & 0.90 & 1.00 \\
\vspace{0.2cm}
   &    & 0.01 & 0.01 & 0.01 & 0.01 & 0.03 & 0.04 & 0.04 \\
   &    & 9.67 & 8.45 & 9.84 & 8.04 & 6.50 & 6.67 & 6.80 \\
$H\alpha$ & 6563 & 2.85 & 2.85 & 2.85 & 2.85 & 2.85 & 2.85 & 2.85 \\
\vspace{0.2cm}
   &    & 0.08 & 0.06 & 0.04 & 0.07 & 0.12 & 0.12 & 0.12 \\
   &    & 3.82 & 3.41 & 3.90 & 3.54 & 3.51 & 4.00 & 3.76 \\
$[N II]$ & 6584 & 1.11 & 1.14 & 1.12 & 1.24 & 1.52 & 1.70 & 1.56 \\
\vspace{0.2cm}
   &    & 0.03 & 0.03 & 0.01 & 0.03 & 0.07 & 0.07 & 0.06 \\
   &    & 1.37 & 0.89 & 0.94 & 0.94 & 1.49 & 1.69 & 1.36 \\
$[S II]$ & 6717 & 0.37 & 0.28 & 0.25 & 0.31 & 0.62 & 0.68 & 0.54 \\
\vspace{0.2cm}
   &    & 0.01 & 0.01 & 0.01 & 0.01 & 0.03 & 0.03 & 0.03 \\
   &    & 0.92 & 0.73 & 0.84 & 0.91 & 1.18 & 1.08 & 1.09 \\
$[S II]$ & 6731 & 0.25 & 0.23 & 0.22 & 0.30 & 0.49 & 0.44 & 0.43 \\
\vspace{0.2cm}
   &    & 0.01 & 0.01 & 0.01 & 0.01 & 0.02 & 0.02 & 0.02 \\
\vspace{0.2cm}
$c$ &    & 1.58 & 1.40 & 1.60 & 1.33 & 1.05 & 1.08 & 1.11 \\
$log\,I\beta$ &    &-13.27 &-12.98 &-12.75 &-13.04 &-13.50 &-13.42 &-13.48 \\
\hline
\end{tabular}

Note - I$\beta$ is the reddening corrected flux in erg cm$^{-2}$ s$^{-1}$, and $\sigma$ is the absolute error of the F$\lambda$/F$\beta$ ratio. Example: Region 3, FH$\alpha$/F$\beta$ = 9.84 $\pm$ 0.04. Distances are given with respect to region 3, being positive toward the NE.

\end{table}

\begin{table}
\footnotesize
\caption{RELATIVE ABUNDANCES AND PHYSICAL CONDITIONS}  \label{abundances}
\begin{tabular}{lccccccc}
\hline
{} & {Region 1} & {Region 2}  & {Region 3}  & {Region 4} & {Region 5} & {Region 6} & {Region 7}\\ 
\vspace{0.2cm}
{Parameter} & {5.1''} & {2.0''}  & {0''}  & {-2''.0} & {-12''.2} & {-14''.2} & {-16''.3}\\ 
\hline
N(O)/N(H) x 10$^4$ & 7.2 & 11.6 & 14.6 & 9.1 & 7.6 & 7.7& 6.6 \\
N(N)/N(H) x 10$^5$ & 4.9 & 8.3 & 10.9 & 9.1 & 7.0 & 9.9 & 8.1 \\
N(N)/N(O) & 0.07 & 0.07 & 0.07 & 0.10 & 0.09 & 0.13 & 0.12 \\
T$_e$  (K) & 7080 & 6140 & 5720 & 6250 & 7030 & 6670 & 6940 \\
N$_e$ (cm$^{-3}$) & 10 & 160 & 270 & 400 & 140 & 10 & 140 \\
\hline
\end{tabular}

Note - Distances are given as in Table ~\ref{fluxes}.

\end{table}

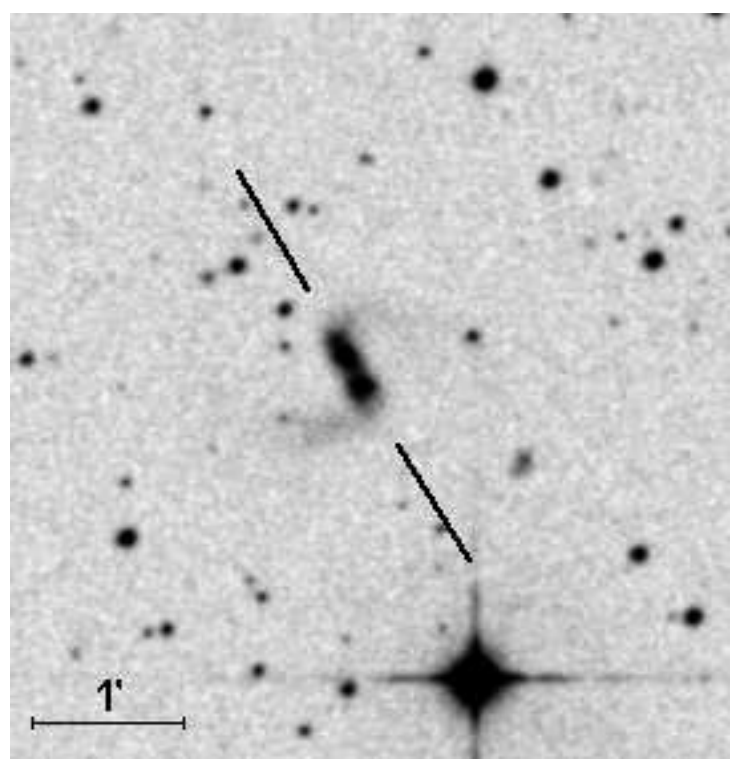
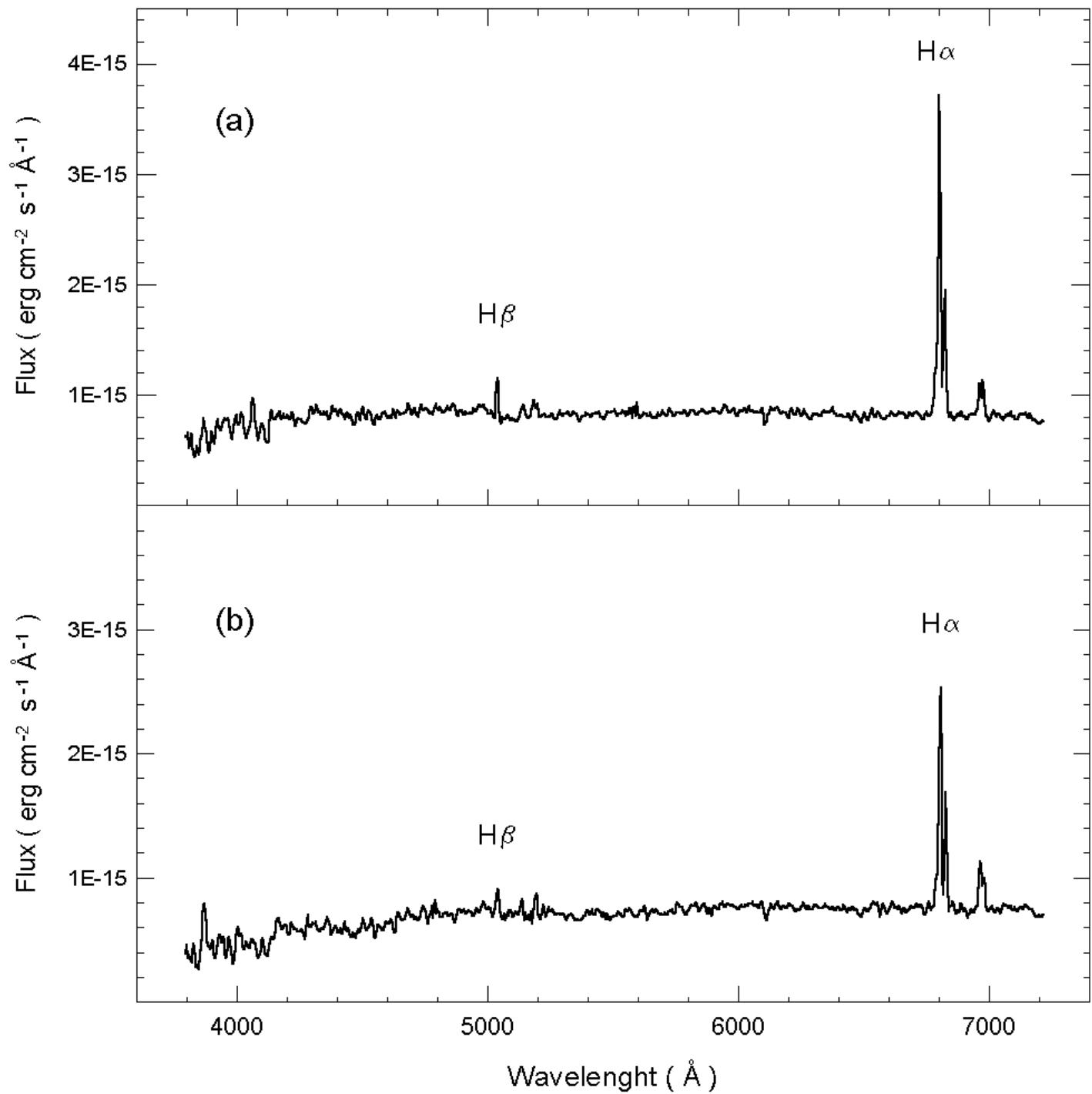
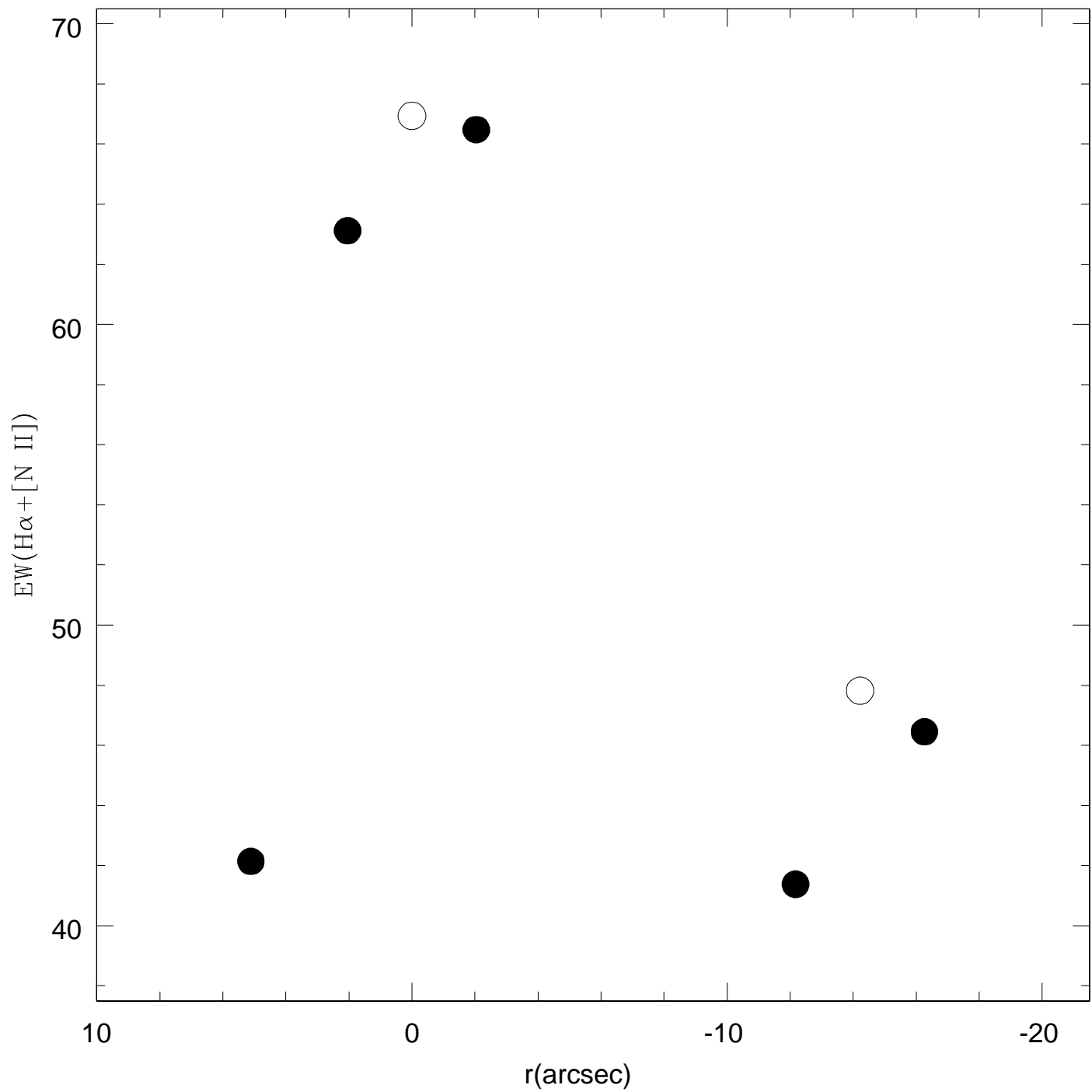
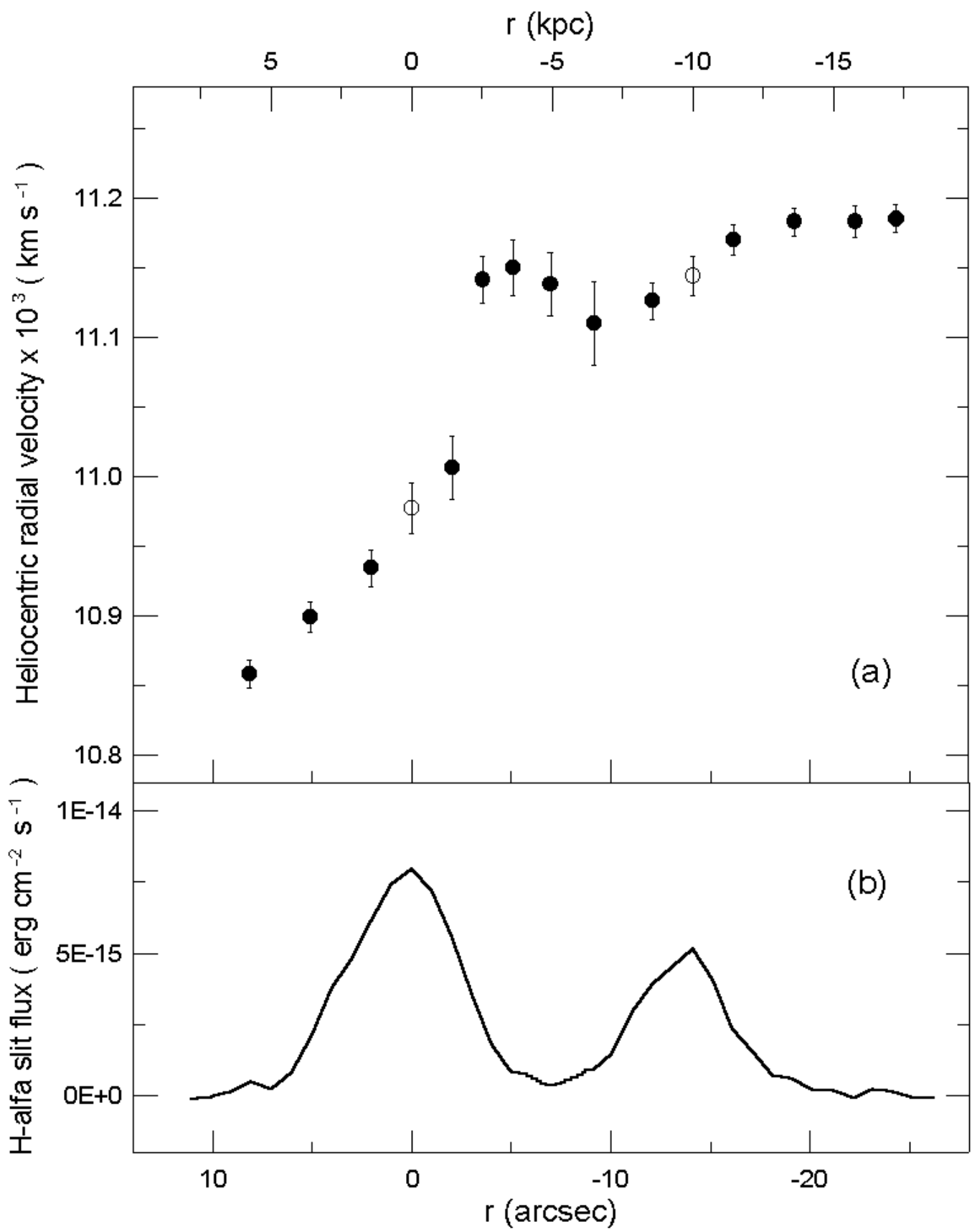
\end{document}